\documentclass{article}
\usepackage{spconf,amsmath,graphicx}
\usepackage{amsfonts}
\usepackage{multirow}
\usepackage{booktabs}

\title{Disentangled Speech Representation Learning for\\One-Shot Cross-lingual Voice Conversion Using $\beta$-VAE}
\name{Hui Lu$^{1,2}$, Disong Wang$^{1}$, Xixin Wu$^{1,2}$, Zhiyong Wu$^{1,3}$, Xunying Liu$^1$, Helen Meng$^{1,2,3}$}
\address{
$^1$The Chinese University of Hong Kong, Hong Kong SAR, China\\
$^2$Centre for Perceptual and Interactive Intelligence, CUHK, Hong Kong SAR, China\\
$^3$Tsinghua-CUHK Joint Research Center for Media Sciences, Technologies and Systems,\\
Shenzhen International Graduate School, Tsinghua University, Shenzhen, China}
\copyrightnotice{978-1-6654-7189-3/22/\$31.00~\copyright2023 IEEE}
\begin{document}
%\ninept
%
\maketitle
\begin{abstract}
We propose an unsupervised learning method to disentangle speech into content representation and speaker identity representation. We apply this method to the challenging one-shot cross-lingual voice conversion task to demonstrate the effectiveness of the disentanglement. Inspired by $\beta$-VAE, we introduce a learning objective that balances between the information captured by the content and speaker representations. In addition, the inductive biases from the architectural design and the training dataset further encourage the desired disentanglement. Both objective and subjective evaluations show the effectiveness of the proposed method in speech disentanglement and in one-shot cross-lingual voice conversion.
\end{abstract}
\begin{keywords}
Speech disentanglement, Voice conversion, cross-lingual, one-shot, unsupervised learning
\end{keywords}
\section{Introduction}
\label{sec:intro}
%Voice conversion (VC) is a technique to convert the speech of the source speaker to make it sound like being uttered by the designated target speaker, while the spoken content remains unchanged \cite{sisman2020overview}.
Voice conversion (VC) is a technique that converts a source speaker's speech to make it sound as though it was uttered by the designated target speaker, while the spoken content remains unchanged \cite{sisman2020overview}.
Cross-lingual VC refers to the scenario where the source and the target speakers do not speak the same language \cite{8683746}.
Cross-lingual VC is more challenging than intra-lingual VC because generally only a monolingual corpus is available from the target speaker, which presents the problem of training-inference mismatch.
Furthermore, one-shot cross-lingual VC is even more challenging, because there is only a single utterance from the target speaker, who speaks a language that is different from the source speaker. We focus on one-shot cross-lingual VC in this paper.

Speech generation embeds different information elements into the signal: content, speaker identity, emotion, language, etc., where the first two elements tend to be more prominent. Generally speaking, language can be considered as part of content. We may regard that a universal phoneme set can cover pronunciation patterns of all languages, and encode different languages in the spoken content.
Following this rationale, we can solve the one-shot cross-lingual VC by disentangling speech into a general content representation and a speaker representation. Then the content representation from the source speech and the speaker representation from the target speech can be combined to generate the speech with the source content and the target speaker identity.

%As is commonly believed that unsupervised learning of disentangled representations is generally impossible, it is important to impose the inductive biases to both the model and the dataset \cite{locatello2019challenging}.
Many methods have been proposed to disentangle the speech into content and speaker representations for one-shot VC.
%, and most of them are based on auto-encoders (AEs) or variational auto-encoders (VAEs) \cite{DBLP:journals/corr/KingmaW13}.
Commonly adopted architectures are the auto-encoder (AE) and variational auto-encoder (VAE) \cite{DBLP:journals/corr/KingmaW13}, with encoder(s) to extract a frame-level feature sequence and a single vector separately from the speech, aiming for the former to encode spoken content, and the latter to encode speaker identity.
This design is based on the fact that the speaker identity is a sequence-level element while the content is a frame-level element. However, as we will discuss in Sec \ref{subsec:architecture}, this alone is not enough to ensure the disentanglement of the content and speaker representations. Many methods have been proposed to further regularize the representation learning to facilitate disentanglement.

AutoVC \cite{DBLP:conf/icml/QianZCYH19} proposes to apply down-sampling and dimension restriction to the content representation to remove the speaker identity. 
VQVC and VQVC+ \cite{DBLP:conf/icassp/WuL20b,DBLP:conf/interspeech/WuCL20} apply vector quantization (VQ) to the content representation to eliminate the speaker information. However, through explicit down-sampling or VQ operations, these methods may hurt the fine-grained temporal information and cause conversion quality degradation.
The mutual information (MI) between the content and the speaker representations has also been imposed as a learning objective \cite{DBLP:conf/iclr/YuanCZHGC21,DBLP:conf/interspeech/WangDYCLM21} to encourage them to be statistically independent, but the estimation of MI is difficult and may make the training process more complex. Other methods apply instance normalization \cite{DBLP:conf/interspeech/ChouL19} and activation regularization \cite{chen2021again} to the content representation, which can induce less information loss and are easier to implement.
%The aforementioned methods are all proposed and evaluated on intra-lingual one-shot VC, it seems that they can be directly adapted to cross-lingual one-shot VC when trained on a multi-lingual speech corpus as they are all unsupervised learning methods.
%However, since the cross-lingual VC presents greater challenge in terms of the need to learn a language-agnostic content representation, it is not clear whether the aforementioned methods can generalize sufficiently well in this case.

Recently, $\beta$-VAE \cite{DBLP:conf/iclr/HigginsMPBGBML17, DBLP:journals/corr/abs-1804-03599} has been proposed as a variant of VAEs for better latent variable disentanglement. With an extra weight parameter $\beta$ being imposed on the Kullback-Leibler (KL) divergence term, $\beta$-VAE restricts the channel capacity of the latent variable to encode the information of the input data. The original $\beta$-VAE is more suitable for static data such as images. In this paper, we propose a variant of $\beta$-VAE that is specifically for disentanglement of sequential data into time-variant and time-invariant representations. Different from $\beta$-VAE which only has one encoder, we adopt two encoders that are respectively designed for content and speaker representation learning. In addition, two weight parameters $\beta_c$ and $\beta_s$ are imposed on the KL divergence terms for content and speaker representations, respectively. We show theoretically that the KL divergence term is an upper bound of the MI between the latent variable and the input data.
Thus, $\beta_c$ and $\beta_s$ function as two gates to restrict the amounts of information of the data that can be captured by the content and speaker representations, respectively.
With proper $\beta_c$ and $\beta_s$ imposed, the information captured by the two representations are complementary to each other, while the content information and speaker identity information are precisely allocated to them, respectively.
% Besides, the architectural designs of the two encoders take into consideration the patterns of the content information and speaker identity information, which also plays an important part in fac.

Compared to existing one-shot VC methods, the proposed method imposes much simpler restrictions on the latent variable (only two weight parameters on the KL divergence terms) which is easier to implement and train.
We do not apply intentionally bottleneck operations such as dimension reduction or down-sampling to the latent variable, thus less information loss is induced. Furthermore, the proposed method can be explained by information theory, which is theoretically more solid.
\section{Related work}
\label{sec:related_work}
Traditional cross-lingual VC methods rely extensively on the phonetic posteriorgrams (PPGs) \cite{8683746,7552917, Zhou2019AMN,zhao2021towards}, which is a speaker-independent content representation explicitly extracted via a speaker-independent automatic speech recognition (SI-ASR) model.
%where a speaker-independent automatic speech recognition (SI-ASR) model is pre-trained to extract the speaker-independent content representation, which is typically referred to as phonetic posteriorgrams (PPGs).
VC is achieved by first extracting PPGs from the source speech, then converting the PPGs into the target speech with a synthesis model, which can be trained only on the target speaker's speech corpus.
%However, when the target speaker does not have multi-lingual speech data, there will be a training-inference mismatch as the synthesis model is trained on the target-language PPGs-speech pairs but applied to convert the PPGs of the source-language speech. Some works propose to train an average synthesis model \cite{8683746, Zhou2019AMN} on the multiple monolingual corpora and then fine-tune the average model on the target speaker's corpus.
%The average model is expected to capture the pronunciation patterns of different languages and can be fine-tuned to the target speaker with the target language.
% Other methods utilize powerful architectures from the text-to-speech (TTS) model as the synthesis model to obtain more accurate PPGs-to-speech mapping \cite{zhao2021towards}.
Though PPGs can be considered as speaker-independent, they are language-dependent, which means that PPGs of one language may not be a good content representation of another language. Besides, since the PPGs extractor is fixed during the training of the synthesis model, it cannot be optimized for all interested languages. In contrast, the proposed method learns to extract the content representation for all interested languages jointly with other parts of the model, thus can in theory learn a more generalized content representation.

For unsupervised disentangled speech representation learning, FHVAE \cite{DBLP:conf/nips/HsuZG17} propose hierarchical prior distributions for VAE to encourage elements of different time scales to be factorized, while the sequence-level representation is further regularized by a discriminative objective. Based on FHVAE, a method is proposed for one-shot cross-lingual VC \cite{Mohammadi2018InvestigationOU}. The method proposed in this paper is different in terms of the restriction imposed, as we emphasize the effects of two weights on the KL divergence terms and no other learning objective is introduced.
Another recently proposed method for one-shot intra-lingual VC \cite{D-DSVAE-VC} also stresses the effects of imposing weights on two KL divergence terms to encourage disentanglement.
However, the proposed method is different from \cite{D-DSVAE-VC} in the following ways: First, the prior distribution for the content representation of \cite{D-DSVAE-VC} is a trainable autoregressive one, while ours is an fixed isotropic Gaussian. Second, as we will show in Section \ref{subsec:obj}, the current paper presents a different derivation (especially for the MI). Besides, it is claimed in \cite{D-DSVAE-VC} that the right choice of the ratio between the two weights can yield the desired disentanglement, while we argue that the absolute values of both weights can also be important for good disentanglement. 
%Though our analysis of the proposed method involves the derivation of MI, it is not estimated explicitly as an leaning objective as done in \cite{DBLP:conf/iclr/YuanCZHGC21,DBLP:conf/interspeech/WangDYCLM21}.
While some methods \cite{DBLP:conf/iclr/YuanCZHGC21,DBLP:conf/interspeech/WangDYCLM21} estimate the MI between different representations as a loss term, the derivation of the MI in this paper only serves as an explanation of the proposed method, we do not need to estimate it numerically.
\section{Method}
\label{sec:method}
%We start with the introduction of the notations.
Let $\mathbf{x}$ be the acoustic feature extracted from speech, and the objective is to find latent variables $\mathbf{z_c}$ and $\mathbf{z_s}$, such that $\mathbf{z_c}$ encodes only spoken content information, while $\mathbf{z_s}$ exclusively contains speaker identity information. With the assumption that $\mathbf{z_c}$ and $\mathbf{z_s}$ are statistically independent, we can derive the objective function as shown in Eqn. (\ref{eqn:elbo}), following the formalization of VAEs. In Eqn. (\ref{eqn:elbo}), $q_\phi(\mathbf{z_c}|\mathbf{x})$ and $q_\phi(\mathbf{z_s}|\mathbf{x})$, are posterior distributions of $\mathbf{z_c}$ and $\mathbf{z_s}$ given $\mathbf{x}$, respectively, parameterized by $\phi$. $p_\theta(\mathbf{x}|\mathbf{z_c}, \mathbf{z_s})$ denotes the conditional distribution of $\mathbf{x}$ given $\mathbf{z_s}$ and $\mathbf{z_c}$ and is parameterized by $\theta$.
$p(\mathbf{z_c})$ and $p(\mathbf{z_s})$ are prior distributions of $\mathbf{z_c}$ and $\mathbf{z_s}$, respectively, and are defined as isotropic Gaussians in this paper, $p_d(\mathbf{x})$ is the data distribution of $\mathbf{x}$.
%When modeled by neural networks, $q_\phi(\mathbf{z_s}|\mathbf{x})$, $q_\phi(\mathbf{z_c}|\mathbf{x})$, and $p_\theta(\mathbf{x}|\mathbf{z_s}, \mathbf{z_c})$ are respectively referred to as the speaker encoder, content encoder, and decoder.
For simplicity, we refer to the three terms in Eqn. (\ref{eqn:elbo}) as the reconstruction loss, content KL, and speaker KL, respectively.
\begin{align}
\label{eqn:elbo}
\mathcal{L}_{\text{vanilla}}&=-\mathbb{E}_{p_d(\mathbf{x})q_\phi(\mathbf{z_c}|\mathbf{x})q_\phi(\mathbf{z_s}|\mathbf{x})}\left[\log p_\theta(\mathbf{x}|\mathbf{z_c},\mathbf{z_s})\right]\nonumber\\
&+\mathbb{E}_{p_d(\mathbf{x})}\left[\text{KL}\left[q_\phi(\mathbf{z_c}|\mathbf{x})\parallel p(\mathbf{z_c})\right]\right]\nonumber\\
&+\mathbb{E}_{p_d(\mathbf{x})}\left[\text{KL}\left[q_\phi(\mathbf{z_s}|\mathbf{x})\parallel p(\mathbf{z_s})\right]\right]
\end{align}

% We start with the introduction of the architectural design that can facilitate the disentanglement of time-variant elements and the time-invariant elements from speech. We will discuss the limitation of the architectural design alone in achieving disentanglement. Then we present the learning objective that can help ensure the disentanglement of the content representation and speaker representation.
The general unsupervised learning model defined above without any other inductive biases cannot ensure the disentanglement, as is commonly believed \cite{DBLP:conf/icml/LocatelloBLRGSB19}.
To achieve the disentanglement of content and speaker representations with the above formalization, our solutions are as follows:
We first introduce the architectural design that facilitate the separation of time-variant and time-invariant elements in Section \ref{subsec:architecture};
In Section \ref{subsec:obj} we present the learning objective that can encourage $\mathbf{z_c}$ and $\mathbf{z_s}$ to be complementary, while exactly the content information and speaker identity information can be assigned separately to them.
The effect of the training dataset is also emphasized in Section \ref{subsec:data_bias}.
%Eqn. (\ref{eqn:elbo}) defines a very general formalization of the proposed model, which is simply a variant of VAEs (with two encoders). As is %commonly believed that unsupervised learning of disentangled representations is in general impossible without inductive biases \cite{DBLP:conf/icml/LocatelloBLRGSB19},
%we introduce the imposed inductive biases in the following subsections, namely, the architectural design, learning objective and dataset bias.
%We argue that all these inductive biases together can largely ensure the learning of disentangled representations of content and speaker.
\subsection{Architectural Design}
\label{subsec:architecture}
%The proposed architectural design is based on the simple observation that the content varies constantly across an utterance while the speaker identity remains the same for the whole utterance.
A basic observation about speech is that the content varies constantly across an utterance while the speaker identity remains the same for the whole utterance. That is, the content is a time-variant feature while speaker identity is a time-invariant one.
This informs us about the structures of the content and speaker representations.
%That is, the content representation should be a time-variant feature, i.e., a sequence of vectors to capture the variations of pronunciations across an utterance. On the other hand, the speaker representation is time-invariant and can thus be designed as a single vector shared by the whole utterance. 
%To this end, we can define the structure of the latent variables.
Suppose that the acoustic feature $\mathbf{x}$ is of the shape $\mathbb{R}^{T\times D_x}$, where $T$ and $D_x$ denote the number of frames and the dimension of $\mathbf{x}$, respectively.
%Then the content latent $\mathbf{z_c}$ and speaker latent $\mathbf{z_s}$ should have the respective shapes of $\mathbb{R}^{T\times D_c}$ and $\mathbb{R}^{D_s}$, where $D_c$ and $D_s$ are the dimensions of $\mathbf{z_c}$ and $\mathbf{z_s}$, respectively.
Then $\mathbf{z_c}$ and $\mathbf{z_s}$ should have the corresponding shapes of $\mathbb{R}^{T\times D_c}$ and $\mathbb{R}^{D_s}$ to separately capture the time-variant content information and time-invariant speaker identity information, where $D_c$ and $D_s$ are the dimensions of $\mathbf{z_c}$ and $\mathbf{z_s}$, respectively.
% It is not hard to find specific neural network architectures for the content encoder and speaker encoder to extract latent variables of the desired structures.
A content encoder transforming the input frame-wisely and a speaker encoder aggregating the whole input sequence into a single vector should thus be adopted.
%For the content encoder, the frame-level latent variable of the same time length as the acoustic feature should be extracted. In contrast, the speaker encoder compresses the whole acoustic feature sequence into a single vector.
% Our implementation details are introduced in Section \ref{sec:impl}.
%This architecture explicitly defines the structure of the $\mathbf{z_c}$ and $\mathbf{z_s}$, which makes it easier for $\mathbf{z_c}$ to capture the time-variant information and $\mathbf{z_s}$ to encode time-invariant information.
Many methods introduced in Section \ref{sec:intro} have adopted similar architectures.

However, this architecture alone cannot guarantee the disentanglement of content and speaker representations.
The reasons are two-fold: First, the architecture cannot ensure that the two representations learned are complementary to each other. Intuitively, the two representations will strive to capture as much information about the input data as possible to reduce the reconstruction loss. Thus, it is possible that the two representations have overlap in the captured information.
%For example, $\mathbf{z_c}$ may also capture the speaker identity information, while $\mathbf{z_s}$ may contain summary information of both speaker identity and spoken content.
% Second, even if we can restrict the two representations to be complementary, the architecture may not precisely separate the content and speaker identity elements.
Second, even if we can restrict the two representations to be complementary (e.g., via bottleneck), then we can reasonably assume that $\mathbf{z_c}$ and $\mathbf{z_s}$ respectively captures complementary time-variant and time-invariant information of speech, but it is not guaranteed that the learned time-variant and time-invariant elements are exactly content and speaker identity.
The reason is that there are many combinations of time-variant and time-invariant information elements other than the targeted one. 
For example, background noise can (mostly) be considered as a time-invariant element, against the time-variant element of clean speech,
but the architecture alone cannot distinguish this pair of elements from the content-speaker identity pair.
%Other examples of time-invariant and time-variant pairs may gender information, against speech with neutral gender.
% The architecture alone is unable to identify which combination of the time-variant and time-invariant elements is the targeted one.
\subsection{Learning Objective}
\label{subsec:obj}
%To tackle the issues along with the architectural design, i.e., to ensure that: 1) the information captured by $\mathbf{z_c}$ and $\mathbf{z_s}$ are complementary and 2) the complementary factors captured are content and speaker identity instead of other factors pair, the key is to restrict the amounts of information of $\mathbf{x}$ that can be captured by $\mathbf{z_c}$ and $\mathbf{z_s}$.
%Aside from architectural design, learning from data must also ensure that:
Aside from the architectural design to facilitate the separation of time-variant and time-invariant feature into $\mathbf{z_c}$ and $\mathbf{z_s}$, we also need to ensure that:
1) the information captured by $\mathbf{z_c}$ and $\mathbf{z_s}$ are complementary and 2) the complementary elements captured are content and speaker identity instead of other element pairs -- and the key is to restrict the amounts of information of $\mathbf{x}$ that can be captured by $\mathbf{z_c}$ and $\mathbf{z_s}$.
To solve issue 1), we can restrict the overall amount of information captured by $\mathbf{z_c}$ and $\mathbf{z_s}$ together to a low level. With no redundant information encoded, the two representations will be compact and contain complementary information.
To further tackle issue 2), it is important to restrict the relative amounts of information captured in the two representations, such that $\mathbf{z_c}$ and $\mathbf{z_s}$ captures exactly the content part and the speaker identity part of information in $\mathbf{x}$, respectively.
% to encourage that the content and speaker elements (instead of other element pairs) are captured
%Intuitively, speaker identity is more complex than the background noise of a utterance, thus the former one should contain more information than the latter one.

The information captured by a latent variable can be quantified as its MI with the data. Following the typical practice by approximating the joint distribution $p(\mathbf{x},\mathbf{z})$ with its variational version $q_\phi(\mathbf{z}|\mathbf{x})p_d(\mathbf{x})$ \cite{chen2018isolating,DBLP:conf/icml/KimM18}, we can derive the variational MI between $\mathbf{x}$ and $\mathbf{z}$ as shown in Eqn. (\ref{eqn:vi_mi}), where $\mathbf{z}$ is a general variable name covering $\mathbf{z_c}$ and $\mathbf{z_s}$.
\begin{align}
\label{eqn:vi_mi}
\mathbb{I}_{v}(\mathbf{x},\mathbf{z})&=\mathbb{E}_{q_\phi(\mathbf{z}|\mathbf{x})p_d(\mathbf{x})}\left[\log{\frac{q_\phi(\mathbf{z}|\mathbf{x})p_d(\mathbf{x})}{p_d(\mathbf{x})q_\phi(\mathbf{z})}}\right]\nonumber\\
&=\mathbb{E}_{q_\phi(\mathbf{z}|\mathbf{x})p_d(\mathbf{x})}\left[\log{\frac{q_\phi(\mathbf{z}|\mathbf{x})}{p(\mathbf{z})}}+\log{\frac{p(\mathbf{z})}{q_\phi(\mathbf{z})}}\right]\nonumber\\
&=\mathbb{E}_{p_d(\mathbf{x})}\left[\text{KL}\left[q_\phi(\mathbf{z}|\mathbf{x})\parallel p(\mathbf{z})\right]\right]\nonumber\\
&-\text{KL}\left[q_\phi(\mathbf{z})\parallel p(\mathbf{z})\right]
\end{align}
Here $q_\phi(\mathbf{z})$ is the marginal distribution of $\mathbf{z}$ and is defined as $q_\phi(\mathbf{z})=\int_{\mathbf{x}}{q_\phi(\mathbf{z}|\mathbf{x})p_d(\mathbf{x})}d\mathbf{x}$. The results in Eqn. (\ref{eqn:vi_mi}) is different from the MI derived in \cite{D-DSVAE-VC}, which claim that $\mathbb{I}_{v}(\mathbf{x},\mathbf{z})=\mathbb{E}_{p_d(\mathbf{x})}\left[\text{KL}\left[q_\phi(\mathbf{z}|\mathbf{x})\parallel p(\mathbf{z})\right]\right]$. This derivation is resulted from approximating the marginal distribution of $\mathbf{z}$ with its prior $p(\mathbf{z})$, which we consider is generally not applicable. 

%Eqn. (\ref{eqn:vi_mi}) denotes that the KL divergence terms in Eqn. (\ref{eqn:elbo}) are closely related to the mutual information between the respective latent variable with the data, where the deviation is the KL divergence between the marginal distribution and the prior distribution of the latent variable.
Eqn. (\ref{eqn:vi_mi}) denotes that $\mathbb{E}_{p_d(\mathbf{x})}\left[\text{KL}\left[q_\phi(\mathbf{z}|\mathbf{x})\parallel p(\mathbf{z})\right]\right]$ is an upper-bound of $\mathbb{I}_{v}(\mathbf{x},\mathbf{z})$. 
In this sense, an approximate method to restrict the amount of information of $x$ captured by the latent variable $\mathbf{z}$ is to restrict the KL divergence term $\mathbb{E}_{p_d(\mathbf{x})}\left[\text{KL}\left[q_\phi(\mathbf{z}|\mathbf{x})\parallel p(\mathbf{z})\right]\right]$.
Though the term $\text{KL}\left[q_\phi(\mathbf{z})\parallel p(\mathbf{z})\right]$ is also penalized as a side-effect, which encourages the marginal distribution of $\mathbf{z}$ to be more like an isotropic Gaussian and does not other harm.
Motivated by this, we introduce two weight parameters $\beta_c$ and $\beta_s$ to the content KL and speaker KL terms, respectively, with the goal to restrict the information that can be captured by $\mathbf{z_c}$ and $\mathbf{z_s}$. The restricted objective function is shown in Eqn. (\ref{eqn:beta_elbo}), while Eqn. (\ref{eqn:beta_elbo_mi}) explicitly denotes the relationship between two KL terms and their corresponding MI terms.
\begin{align}
\mathcal{L}_{\beta}&=-\mathbb{E}_{p_d(\mathbf{x})q_\phi(\mathbf{z_c}|\mathbf{x})q_\phi(\mathbf{z_s}|\mathbf{x})}\left[\log p_\theta(\mathbf{x}|\mathbf{z_c},\mathbf{z_s})\right]\nonumber\\
&+\beta_c\cdot\mathbb{E}_{p_d(\mathbf{x})}\left[\text{KL}\left[q_\phi(\mathbf{z_c}|\mathbf{x})\parallel p(\mathbf{z_c})\right]\right]\nonumber\\
&+\beta_s\cdot\mathbb{E}_{p_d(\mathbf{x})}\left[\text{KL}\left[q_\phi(\mathbf{z_s}|\mathbf{x})\parallel p(\mathbf{z_s})\right]\right]\label{eqn:beta_elbo}\\
&=-\mathbb{E}_{p_d(\mathbf{x})q_\phi(\mathbf{z_c}|\mathbf{x})q_\phi(\mathbf{z_s}|\mathbf{x})}\left[\log p_\theta(\mathbf{x}|\mathbf{z_c},\mathbf{z_s})\right]\nonumber\\
&+\beta_c\cdot\left[\mathbb{I}_{v}(\mathbf{x},\mathbf{z_c})+\text{KL}\left[q_\phi(\mathbf{z_c})\parallel p(\mathbf{z_c})\right]\right]\nonumber\\
&+\beta_s\cdot\left[\mathbb{I}_{v}(\mathbf{x},\mathbf{z_s})+\text{KL}\left[q_\phi(\mathbf{z_s})\parallel p(\mathbf{z_s})\right]\right] \label{eqn:beta_elbo_mi}
\end{align}

While the reconstruction loss in Eqn. (\ref{eqn:beta_elbo}) encourages both $\mathbf{z_c}$ and $\mathbf{z_s}$ to capture as much information about $\mathbf{x}$ as possible, 
%the two KL divergence terms penalize the amounts of information that is captured by them, as indicated by Eqn. (\ref{eqn:beta_elbo_mi}).
the weight parameters $\beta_c$ and $\beta_s$ act as two gates to control the amounts of information of $\mathbf{x}$ going through $\mathbf{z_c}$ and $\mathbf{z_s}$, as indicated by Eqn. (\ref{eqn:beta_elbo_mi}).
% Ideally, with properly chosen $\beta_c$ and $\beta_s$, $\mathbf{z_c}$ and $\mathbf{z_s}$ should capture exact the information of content and speaker identity.
The learning objective (\ref{eqn:beta_elbo}) can help solve the problems of the architectural design in the following ways:
By choosing the proper absolute values of both $\beta_c$ and $\beta_s$, the information captured by $\mathbf{z_c}$ and $\mathbf{z_s}$ together will be compact and thus the information contained in each latent variable will be complementary.
%On the other hand, by setting proper relative values of $\beta_c$ and $\beta_s$, the content and speaker identity information can be exactly allocated to $\mathbf{z_c}$ and $\mathbf{z_s}$, respectively.
On the other hand, by tuning relative values of $\beta_c$ and $\beta_s$, the amounts of information in $\mathbf{x}$ allocated to $\mathbf{z_c}$ and $\mathbf{z_s}$ will change accordingly. In this case, we can find the optimal parameters pair that yields exactly the separation of content and speaker identity.

Note that if we let $\beta_c=\beta_s$ and combine $\mathbf{z_c}$ and $\mathbf{z_s}$ together, Eqn. (\ref{eqn:beta_elbo}) will become exactly the learning objective of the standard $\beta$-VAE \cite{DBLP:conf/iclr/HigginsMPBGBML17}. In this sense, Eqn. (\ref{eqn:beta_elbo}) can be considered as a generalized form of $\beta$-VAE.
\subsection{Dataset bias}
\label{subsec:data_bias}
The dataset used to train the model is also a very important inductive bias to facilitate the learning of disentangled representations. One consideration is that the two information elements to be disentangled should vary independently in the dataset. Besides, other elements of variation that may interfere the interested elements should not appear in the dataset. For example, when disentangling the content and speaker identity elements, too much variation in emotions may disturb the learning of speaker identity element since they are both sequence-level elements. In our case, we combine two monolingual corpora as the training dataset, in which the speaker and general content are the main variations. More details are given in Section \ref{subsec:exp_data}.
\section{Implementation}
\label{sec:impl}
The proposed model consists of a speaker encoder, a content encoder, and a decoder, to respectively model $q_\phi(\mathbf{z_s}|\mathbf{x})$, $q_\phi(\mathbf{z_c}|\mathbf{x})$, and $p_\theta(\mathbf{x}|\mathbf{z_c},\mathbf{z_s})$ defined in Section \ref{sec:method}. The structure of each component as well as other implementation details are described as follows.
% \begin{itemize}

\textbf{Speaker encoder}: The speaker encoder consists of one fully-connected (FC) layer with 256 hidden units, whose output is activated by ReLU \cite{agarap2018deep} and fed into 4 down-sampling convolutional blocks.
Each down-sampling convolutional block includes two 1D convolution (Conv) layers with 256 filters and ReLU activations, while the output of the second Conv layer is down-sampled by a factor of 2 using average pooling along the time axis.
A residual connection is adopted in each convolutional block to connect the input and the output.
The kernel sizes of Conv layers in the 4 convolutional blocks are 3, 3, 5 and 5, respectively. The output of convolutional blocks is further aggregated through global average-pooling and fed into a fully connected layer to get the 128-dimensional mean and variance vector of the speaker posterior.

\textbf{Content encoder}: The content encoder consists of two 1D Conv layers each with 256 hidden units and a kernel size of 3.
Dropout \cite{srivastava2014dropout} with the drop rate of 0.2 is embedded after each Conv layer.
Two self-attention blocks \cite{vaswani2017attention} with 256 hidden units, 4 attention heads, and 1024 feed-forward network hidden dimensions are further stacked. The output of self-attention blocks is projected to the 128-dimensional mean and the variance for each frame.
%During the training phase, the content representation is sampled frame-wisely from the predicted distribution, while for inference the mean sequence can be used as the content representation. 

\textbf{Decoder}: The decoder takes in the concatenation of the speaker representation and the content representation to reconstruct the acoustic feature. Conv layers and self-attention blocks with the same configurations as those in the content encoder are first adopted. An FC layer is then used to predict the acoustic feature. We further apply a PostNet \cite{shen2018natural} module which consists of 5 layers of 1D Conv with the kernel size of 5, to predict the residual of the acoustic feature. Dropout with probability 0.2 is used to regularize the PostNet.
Both acoustic features predicted before and after the PostNet are taken out for the loss computation.

\textbf{Loss function}: While Eqn. (\ref{eqn:beta_elbo}) defines the theoretical form of the learning objective, which consists the negative log-likelihood of the conditional distribution of the acoustic feature as the reconstruction loss. In practice, we adopt the sum of the mean-squared error (MSE) loss and mean-absolute error (MAE) as the reconstruction loss.
Since the conditional distribution defined by our reconstruction loss cannot be explicitly expressed, the $\beta_c$ and $\beta_s$ values are not normalized \cite{DBLP:conf/iclr/HigginsMPBGBML17} with respect to the standard form of VAE.

\textbf{Training and inference}: During training, the inputs to the speaker encoder and content encoder are from the same utterance, while the input to the speaker encoder is firstly segmented and shuffled along the time axis before being fed into the following network. This operation can help avoid the content information being leaked into the speaker representation.
The speaker latent and the content latent are sampled though re-parameterization during training, while only the mean vector and mean sequence of the two latent are used during inference. For model optimization, we use Adam optimizer \cite{Kingma2015AdamAM} with $\beta_1=0.9$ and $\beta_2=0.999$ and $\epsilon=10^{-7}$, the learning rate is fixed at $1.25\times10^{-4}$. The training batch size is set as 32. During inference, the speech from the target speaker is fed into the speaker encoder to obtain the speaker representation, which is combined with content extracted from the source speech in the decoder to generated the converted speech.

\textbf{Hyper-parameters tuning}: The two most important hyper-parameters for the proposed model are $\beta_c$ and $\beta_s$, which directly determines the disentanglement performance. To be honest, the tuning of these two parameters relies largely on trial and error, i.e., one can find the best setting through grid search. But we do notice one simple trick that can help find the good parameters faster. One can start with a relatively large $\beta_c$ (e.g., 0.1, if the same loss function as ours is used) and a small $\beta_s$ (e.g., $10^{-3}$), this can generally yield the content and speaker disentanglement, but the generation quality may not be that good since the large $\beta_c$ causes too much information loss of the content. Then we can decrease $\beta_c$ gradually to find the value that yields also good generation quality. At the same time, $\beta_s$ should first be decreased proportionally with $\beta_c$ and then separately tuned.
% \end{itemize}
\section{Experiments}
\label{sec:exp}
%The goal of the experiments is to demonstrate: 1) the effectiveness of the proposed method in learning disentangled content and speaker representations of speech, and 2) the effectiveness of the proposed method in performing one-shot cross-lingual VC.
\subsection{Dataset}
\label{subsec:exp_data}
We combine two openly available corpora, which are VCTK \cite{Yamagishi2019CSTRVC} and AISHELL-3 \cite{shi21c_interspeech} together for training and evaluation of the proposed model. VCTK contains 110 speakers' English speech data while AISHELL-3 consists of speech uttered by 218 native Mandarin speakers.
% For VCTK, the numbers of speakers used for training, validataion and testing are respectively 88, 10 and 10, while these numbers for AISHELL-3 are 116, 15 and 16.
88 speakers' data from VCTK and 116 speakers' data from AISHELL-3 are combined as a bilingual training set.
Another 20 speakers' data from VCTK are evenly split for validation and testing, while for AISHELL-3 15 and 16 other speakers' data are used so.
%10 speakers' data from VCTK and 15 speakers' data from AISHELL-3 are used as the validation set. Another 10 speakers' data from VCTK and 16 speakers' data from AISHELL-3 are combined as the test set.
Note that there are no common speakers among different splits of the dataset. We down-sample all speech into 16kHz and extract 80-dimensional logarithm Mel-Spectrograms with 200ms window length and 50ms window shift as the feature.
\subsection{Evaluation on disentanglement}
\label{subsec:eval_dis}
Following prior works \cite{DBLP:conf/nips/HsuZG17,D-DSVAE-VC,DBLP:conf/icml/LiM18a}, we conduct speaker verification (SV) on the learned content and speaker representations and report the equal error rate (EER) as a metric of the disentanglement. Intuitively, a high EER produced by $\mathbf{z_c}$ and a low EER yielded by the $\mathbf{z_s}$ can denote a good disentanglement of content and speaker representations. To compute the EER, we randomly select 4 utterances from each speaker in the test set as the enrolled samples, which are used to compute the speaker embedding (by averaging their $\mathbf{z_c}$'s or $\mathbf{z_c}$'s). The remaining utterances of the speaker are used as the positive trials while all other speakers' utterances are negative trials. Cosine similarity is computed as the score. The EERs are evaluated separately for English and Mandarin test sets.

As discussed in Section \ref{subsec:obj}, the weight parameters $\beta_c$ and $\beta_s$ restrict the respective amounts of information contained in $\mathbf{z_c}$ and $\mathbf{z_s}$, and thus are very important for disentanglement. We show EERs with respect to $\mathbf{z_c}$ and $\mathbf{z_s}$ for both English (EN) and Mandarin (CN) test set in Table \ref{tab:eer}, where $\beta_c$ varies among $\{10^{-3}, 10^{-2}\}$ and $\beta_s$ is taken from $\{10^{-5}, 10^{-4}, 10^{-3}\}$.

We can first observe the effect of decreasing the absolute values of both $\beta_c$ and $\beta_s$ by comparing the results yielded by cases $(\beta_c, \beta_s)=(10^{-2}, 10^{-4})$ and $(\beta_s, \beta_c)=(10^{-3}, 10^{-5})$. While the ratios between $\beta_c$ and $\beta_s$ for both cases are 100, they produce quite different representations in terms of disentanglement. As we can observe from results of English test set, the EERs computed using $\mathbf{z_c}$ and $\mathbf{z_s}$ are $0.369$ and $0.115$ respectively for the former case, while those values become $0.198$ and $0.069$ for the latter case. While the former case shows more desirable disentanglement, the second case yields a $\mathbf{z_c}$ with too much speaker information. This is because with restrictions that are too loose on both $\mathbf{z_c}$ and $\mathbf{z_s}$ when setting $(\beta_s, \beta_c)=(10^{-3}, 10^{-5})$, it cannot be ensured that the information captured by $\mathbf{z_c}$ and $\mathbf{z_s}$ are complementary. Thus, different from \cite{D-DSVAE-VC} which claims that a proper ratio $\frac{\beta_c}{\beta_s}$ can induce the desired disentanglement, we argue that it is also important to first set proper absolute values for $\beta_c$ and $\beta_s$, which restricts $\mathbf{z_c}$ and $\mathbf{z_s}$ to be more complementary.

Furthermore, setting proper relative values of $\beta_c$ and $\beta_s$ is also important for disentanglement.
As can be observed from Table \ref{tab:eer}, increasing the value of $\beta_c$ causes significant increases of EERs for both EN and CN content representations and all $\beta_s$ values, denoting the reduction of the speaker information captured by $\mathbf{z_c}$. On the other hand, the increase of $\beta_s$ puts more penalization on the speaker representation and thus allows more speaker information to leak into $\mathbf{z_c}$, which is indicated by the overall declining trend of EERs for $\mathbf{z_c}$ and the opposite trend for $\mathbf{z_s}$ horizontally.
Meanwhile, larger $\beta_c$ ($10^{-2}$) produces a stabler speaker-independent $\mathbf{z_c}$ as the changes in EERs caused by the increasing $\beta_s$ are limited.
%Meanwhile, the respective changes caused by increasing $\beta_c$ in EERs on $\mathbf{z_s}$ are not so obvious compared to those on $\mathbf{z_c}$. The reason may be that $\mathbf{z_c}$ is more complex in structure and can bear 

% It is claimed in \cite{D-DSVAE-VC} that a proper ratio $\frac{\beta_c}{\beta_s}$ can induce the desired disentanglement. However, as we can observe in Table \ref{tab:eer}, though the ratios of cases $(\beta_c, \beta_s)=(1e-2, 1e-4)$ and $(\beta_s, \beta_c)=(1e-3, 1e-5)$ are both 100, the disentanglement results they yield are different. For English, the EERs computed using $\mathbf{z_c}$ and $\mathbf{z_s}$ are $0.369$ and $0.115$ respectively for the former case, while those values become $0.198$ and $0.069$ for the latter case. While the former case shows more desirable disentanglement, the second case yields a $\mathbf{z_c}$ with too much speaker information. This is because with restrictions that are too loose on both $\mathbf{z_c}$ and $\mathbf{z_s}$ when setting $(\beta_s, \beta_c)=(1e-3, 1e-5)$, it cannot be ensured that the information captured by $\mathbf{z_c}$ and $\mathbf{z_s}$ are complementary. Thus, we argue that it is also important to set proper absolute values for $\beta_c$ and $\beta_s$.
\begin{table}[th]
  \caption{SV results on content and speaker representations}
  \label{tab:eer}
  \centering
  \resizebox{0.4\textwidth}{!}{
  \begin{tabular}{ccccc}
    \toprule
    Rep. & &                      $\beta_s=10^{-5}$ & $\beta_s=10^{-4}$ & $\beta_s=10^{-3}$\\
    \midrule
    \multirow{2}{*}{$\mathbf{z_c}$ (EN) $\uparrow$} & $\beta_c=10^{-3}$ & 0.198 & 0.179 & 0.151~~~ \\
                                & $\beta_c=10^{-2}$ & 0.366 & 0.369 & 0.319~~~\\
		\midrule
		\multirow{2}{*}{$\mathbf{z_c}$ (CN) $\uparrow$} & $\beta_c=10^{-3}$ & 0.253 & 0.223  & 0.155~~~ \\
                                & $\beta_c=10^{-2}$ & 0.379 & 0.375 &0.328~~~\\
		\midrule
    \multirow{2}{*}{$\mathbf{z_s}$ (EN) $\downarrow$} & $\beta_c=10^{-3}$ & 0.069 & 0.131 & 0.372~~~ \\
                                & $\beta_c=10^{-2}$ & 0.075 & 0.115 & 0.357~~~\\
		\midrule
		\multirow{2}{*}{$\mathbf{z_s}$ (CN) $\downarrow$} & $\beta_c=10^{-3}$ & 0.061 & 0.105 & 0.273~~~ \\
                                & $\beta_c=10^{-2}$ & 0.043 & 0.113 & 0.285~~~ \\
    \bottomrule
  \end{tabular}}
\end{table}
\vspace*{-\baselineskip}
\subsection{Evaluation on one-shot VC}
\label{subsec:eval_vc}
% The performance of VC is also an important indicator of content and speaker disentanglement.
We conduct subjective evaluations on both intra-lingual and cross-lingual VC. We select 4 Mandarin speakers and 4 English speakers, both consisting of 2 male speakers and 2 female speakers. One utterance of each speaker is randomly chosen as the source and also the reference speech. All utterances are converted to all other speakers, thus in total we obtain 56 converted samples, while 24 of them are intra-lingual (EN2EN and CN2CN) and 32 are cross-lingual cases (EN2CN and CN2EN).
%12 native Mandarin and L2 English speakers
Twelve subjects are asked to score these converted samples based on their naturalness and speaker similarity. 5-scale mean opinion score (MOS) is applied for both speech naturalness and speaker similarity.

We adopt two competitive unsupervised one-shot VC models AdIN-VC \cite{DBLP:conf/interspeech/ChouL19} and VQMIVC \cite{DBLP:conf/interspeech/WangDYCLM21} as our baselines, while Hifi-GAN \cite{kong2020hifi} is used as the vocoder. We denote the proposed model as $\beta$-VAEVC. We train the two baseline models as well as the vocoder on the same acoustic features and training set as ours. The copy-synthesized speech of the source speech is included in the naturalness evaluation, while those of randomly selected speech from the target speakers are scored in the speaker similarity evaluation. Some samples are available on \textit{https://beta-vaevc.github.io}.

We first show the EERs obtained by content and speaker representations of three compared models in Table \ref{tab:eer_all_models}. For $\beta$-VAEVC the $\beta_c$ and $\beta_s$ are set as $3\times 10^{-3}$ and $10^{-7}$ respectively.
However, as shown in Table \ref{tab:eer_all_models}, this setting does not yield the best disentanglement results in terms of EERs.
As $\mathbf{z_c}$ of $\beta$-VAEVC seems to contain more speaker information than two baselines.
Though we can achieve more speaker-independent $\mathbf{z_c}$ for $\beta$-VAEVC via further increasing $\beta_c$, we find it can also decrease the generation quality.
% This trade-off is also indicated by Eqn. (\ref{eqn:beta_elbo}), where larger restrictions on $\mathbf{z_c}$ and $\mathbf{z_s}$ can cause increase of the reconstruction loss.
% Though larger $\beta_c$ on content KL will make it more speaker-independent, as shown in Table \ref{tab:eer}, it also leads to more information loss, thus can degrade the generation quality. This can be considered as a trade-off between disentanglement and generation quality. 
Thus, we choose the parameter setting that works better on the validation set in terms of speech generation quality.
\begin{table}[th]
  \caption{SV results of three compared models}
  \label{tab:eer_all_models}
  \centering
  \resizebox{0.4\textwidth}{!}{
  \begin{tabular}{c c c c c}
    \toprule
    Model     & $\mathbf{z_c}$ (EN) $\uparrow$   & $\mathbf{z_s}$ (EN) $\downarrow$   & $\mathbf{z_c}$ (CN) $\uparrow$    & $\mathbf{z_s}$ (CN) $\downarrow$ \\
    \midrule
    AdIN-VC	        &0.373           &0.065          &0.371          &0.074~~~          \\
    VQMIVC	        &\textbf{0.398}  &0.106	         &\textbf{0.376} &0.091~~~          \\
    $\beta$-VAEVC	&0.279	         &\textbf{0.054} &0.322          &\textbf{0.053}~~~ \\
    \bottomrule
  \end{tabular}}
\end{table}
The naturalness and similarity MOS results are shown in Table \ref{tab:naturalness_mos} and Table \ref{tab:similarity_mos}, respectively. We can observe that the proposed method achieves overall both better naturalness and speaker similarity than the two baselines. While AdIN-VC works comparably well for English intra-lingual conversion, its performance is much worse when it comes to Mandarin intra-lingual conversion and two directions of cross-lingual conversion.
While VQMIVC archives overall good conversion naturalness, the speaker similarity performance is not as satisfying as that for speech naturalness, especially for cross-lingual cases.
\begin{table}[th]
  \caption{Speech naturalness MOS results ($\pm95\%$ CI)}
  \label{tab:naturalness_mos}
  \centering
	\resizebox{0.48\textwidth}{!}{
  \begin{tabular}{c c c c c c}
    \toprule
    Model     & EN2EN      & EN2CN       & CN2CN     & CN2EN      & Overall\\
    \midrule
    Hifi-GAN	 & 4.32±0.08	& 4.32±0.08  & 4.30±0.08 &	4.30±0.08	& 4.31±0.06~~~          \\
    \midrule
    AdIN-VC  &	3.41±0.15 & 2.85±0.14	  & 2.94±0.17  & 2.74±0.15 & 2.96±0.08~~~               \\
    VQMIVC	&3.56±0.15	&3.15±0.12	&3.27±0.13	&3.18±0.14	&3.27±0.07~~~               \\
    $\beta$-VAEVC	&3.71±0.14	&\textbf{3.53±0.13}	&\textbf{3.71±0.14}	&3.35±0.15	&\textbf{3.56±0.07}~~~               \\
    \bottomrule
  \end{tabular}}
\end{table}
\vspace*{-\baselineskip}
\begin{table}[th]
  \caption{Speaker similarity MOS results ($\pm95\%$ CI)}
  \label{tab:similarity_mos}
  \centering
	\resizebox{0.48\textwidth}{!}{
  \begin{tabular}{c c c c c c}
    \toprule
    Model     & EN2EN     & EN2CN       & CN2CN     & CN2EN      & Overall\\
    \midrule
    Hifi-GAN	&4.30±0.13	&4.46±0.10	&4.46±0.10	&4.30±0.13	&4.36±0.08~~~          \\
    \midrule
    AdIN-VC	&3.36±0.23	&2.83±0.18	&2.75±0.24	&2.91±0.21	&2.95±0.11~~~               \\
    VQMIVC	&3.04±0.25	&2.60±0.19	&3.26±0.22	&2.55±0.20	&2.82±0.11~~~               \\
    $\beta$-VAEVC	&3.54±0.19	&3.00±0.16	&3.32±0.24	&3.31±0.18	&\textbf{3.27±0.10}~~~               \\
    \bottomrule
  \end{tabular}}
\end{table}
% (WER) for English and  (CER) for Mandarin

In addition, we utilize the transcription error obtained by open-source pre-trained ASR models \cite{kuchaiev2019nemo} as indicators of the conversion intelligibility. We conduct both intra-lingual and cross-lingual VC on the whole test set, that is, all utterances are converted to all the other speakers. There are in total 42,340 converted utterances for English intra-lingual VC, 77,920 for Mandarin intra-lingual VC, 67,744 for English-to-Mandarin conversion and 48,700 for Mandarin-to-English conversion. Pre-trained English and Mandarin ASR models are used to transcribe the corresponding utterances, then the word error rate (WER) and character error rate (CER) are computed respectively for English and Mandarin.
The recognition results on re-synthesized utterances of all samples in the test set by Hifi-GAN are also included as reference. The results are shown in Table \ref{tab:wer_cer}.
The proposed model surpasses baseline models by large margins for two intra-lingual VC cases and the Mandarin-to-English conversion case, and shows comparable performance with other two methods for the English-to-Mandarin conversion. Besides, we notice that VQMIVC achieves better WER / CER than AdIN-VC for cross-lingual conversion, while the latter model is better for intra-lingual cases.

Though all compared models can realize cross-lingual VC, we can see that there are gaps between the performance of intra-lingual VC and cross-lingual VC for all metrics, as shown in Table \ref{tab:naturalness_mos}, \ref{tab:similarity_mos} and \ref{tab:wer_cer}.
This is reasonable since the inputs to the content encoder and speaker encoder are from the different domain for cross-lingual VC cases, which is not the case during training. Besides, the difference in the recording conditions of two corpora VCTK and AISHELL-3 can make the train-inference mismatch issue even worse. We will tackle this problem in our future work.
\vspace*{-\baselineskip}
% Part of the reason may be that the recording conditions of two corpora AISHELL-3 and VCTK are not the same, this may introduce the out-of-distribution inputs during inference, i.e., the inputs to the content encoder and speaker encoder are not compatible with each other. This can cause the conversion performance degradation for both naturalness and speaker similarity. For the WER / CER test, the out-of-distribution issue caused by the cross-lingual conversion can be more severe.
\begin{table}[th]
  \caption{Speech recognition error results}
  \label{tab:wer_cer}
  \centering
	\resizebox{0.4\textwidth}{!}{
  \begin{tabular}{c c c c c}
    \toprule
    Model     & EN2EN      & EN2CN       & CN2CN     & CN2EN \\
    \midrule
    Hifi-GAN  &\multicolumn{2}{c}{5.11\%} &\multicolumn{2}{c}{2.70\%}~~~          \\
    \midrule
    AdIN-VC	&30.61\%	&45.88\%	&23.01\%	&51.31\%~~~               \\
    VQMIVC	&35.09\%	&\textbf{43.60\%}	&33.37\%	&44.70\%~~~               \\
    $\beta$-VAEVC	&\textbf{23.41\%}	&46.33\%	&\textbf{10.58\%}	&\textbf{32.24\%}~~~               \\
    \bottomrule
  \end{tabular}}
\end{table}
\vspace*{-\baselineskip}
\section{Conclusion}
\label{sec:conclusion}
We propose a method to disentangle speech into content and speaker representations, which can be applied to the challenging task of one-shot cross-lingual VC.
Our method is based on a VAE with two encoders to extract speaker and content representations respectively. With the speaker encoder compressing the whole speech into a single vector and the content encoder extracting the frame-level representation out of speech, the time-variant and time-invariant elements of speech can be more easily separated into two representations. Furthermore, inspired by $\beta$-VAE, we propose a learning objective that incorporates two weight parameters to restrict the amount of information that can be captured by the two representations. With proper weight parameters imposed, the disentanglement can be ensured to be with respect to content and speaker information. We apply the proposed method to one-shot cross-lingual VC, through which we show the effectiveness of the proposed method in achieving content and speaker disentanglement. 
\section{ACKNOWLEDGMENTS}
\label{sec:ack}
This research is supported by the Centre for Perceptual and Interactive Intelligence, a CUHK InnoCentre.
\bibliographystyle{IEEEbib}
\bibliography{strings,refs}

\end{document}